\documentclass[apj]{emulateapj} 

\shortauthors{KILIC ET AL.} 
\shorttitle{The Ages of the Disk and the Halo}

\begin{document} 
\title{The Ages of the Thin Disk, Thick Disk, and the Halo from Nearby White Dwarfs}

\author{
Mukremin Kilic\altaffilmark{1}, 
Jeffrey A. Munn\altaffilmark{2},
Hugh C. Harris\altaffilmark{2},
Ted von Hippel\altaffilmark{3,4},
James W. Liebert\altaffilmark{5},
Kurtis A. Williams\altaffilmark{6},
Elizabeth Jeffery\altaffilmark{7},
Steven DeGennaro\altaffilmark{8}
}
\altaffiltext{1}{Homer L. Dodge Department of Physics and Astronomy, University of Oklahoma, 440 W. Brooks St., Norman, OK, 73019}
\altaffiltext{2}{US Naval Observatory, Flagstaff Station, 10391 W. Naval Observatory Road, Flagstaff, AZ 86005}
\altaffiltext{3}{Center for Space and Atmospheric Research, Embry-Riddle Aeronautical University, Daytona Beach, FL 32114}
\altaffiltext{4}{Max Planck Institute for Astronomy, K\"onigstuhl 17, 69117 Heidelberg, Germany}
\altaffiltext{5}{University of Arizona, Steward Observatory, Tucson, AZ 85721}
\altaffiltext{6}{Department of Physics and Astronomy, Texas A\&M University-Commerce, P.O. Box 3011, Commerce, TX 75429}
\altaffiltext{7}{Department of Physics and Astronomy, Brigham Young University, N283 ESC, Provo, UT 84602}
\altaffiltext{8}{Department of Astronomy, University of Texas at Austin, 1 University Station C1400, Austin, TX 78712}

\begin{abstract}

We present a detailed analysis of the white dwarf luminosity functions derived from
the local 40 pc sample and the deep proper motion catalog of \citet{munn14,munn17}.
Many of the previous studies ignored the contribution of thick disk white dwarfs to
the Galactic disk luminosity function, which results in an erronous age measurement.
We demonstrate that the ratio of thick/thin disk white dwarfs is roughly 20\% in the
local sample. Simultaneously fitting for both disk components, we derive ages of
6.8-7.0 Gyr for the thin disk and 8.7 $\pm$ 0.1 Gyr for the thick disk from the
local 40 pc sample. Similarly, we derive ages of 7.4-8.2 Gyr for the thin disk 
and 9.5-9.9 Gyr for the thick disk from the deep proper motion catalog, which shows
no evidence of a deviation from a constant star formation rate in the past 2.5 Gyr.
We constrain the time difference between the onset of star formation in the thin disk
and the thick disk to be $1.6^{+0.3}_{-0.4}$ Gyr. The faint end of the luminosity
function for the halo white dwarfs is less constrained, resulting in an age
estimate of $12.5^{+1.4}_{-3.4}$ Gyr for the Galactic inner halo. This is the first
time ages for all three major components of the Galaxy are obtained from a sample
of field white dwarfs that is large enough to contain significant numbers of disk
and halo objects. The resultant ages agree reasonably well with the age estimates
for the oldest open and globular clusters.

\end{abstract}


\section{Introduction}

There are a variety of methods for measuring the ages of stars and the populations
that they reside in. Traditionally the best age measurements have come from fits to
the main-sequence turn-offs in open and globular clusters, which reveal ages of up
to $\sim$12 Gyr for the oldest globular clusters. The same technique can be applied 
to field stars if accurate trigonometric parallaxes are available. For example,
using Hipparcos parallaxes of turn-off field stars and subgiants, \citet{liu00}
and \citet{sandage03} derive metallicity-dependent solar neighborhood disk ages of 
7.5 - 7.9 $\pm$ 0.7 Gyr. 

Gyrochronology can be used to age-date clusters as old
as 4 Gyr with a precision of $\sim$0.7 Gyr \citep{barnes16}. However,
\citet{angus15} find unexpected deviations from the predicted age versus rotation
period relation and they demonstrate that the age uncertainties are significantly higher
for field stars.

Nucleocosmochronometry, the use of thorium and uranium to infer
ages of metal-poor halo stars, provide ages with an uncertainty of 2-3 Gyr for
13 Gyr old halo stars \citep{sneden03,frebel07}. There is no single method that
works well for a broad range of stellar types or ages \citep{soderblom10}, and
each method has its own theoretical (e.g., incomplete treatments of convection and
rotation for main-sequence stars) and observational (e.g., small sample size,
uncertainties in abundances, distances, and reddening) problems that limit the age
precision.

White dwarfs offer an independent technique for measuring stellar population ages.
The surface temperature and luminosity of a white dwarf is primarily a function
of its age, and secondarily a function of its mass, interior and surface composition.
\citet{schwarzschild58} estimated a luminosity of $10^{-4} L_{\odot}$ for an 8 Gyr 
old white dwarf. \citet{winget87} and \citet{liebert88} demonstrate that the lack of
stars fainter than this luminosity in the Solar neighborhood is due to the
finite age of the Galactic disk, which \citet{leggett98} constrain to be $8 \pm 1.5$ Gyr
based on 43 stars.
\citet{hansen04,hansen07,hansen13} extended white dwarf cosmochronology to the halo
by using {\em Hubble Space Telescope} observations of the globular clusters
M4, NGC 6397, and 47 Tuc, and derived ages of $\geq$10.3, 11.5 $\pm$ 0.5, and 9.9 $\pm$ 0.7 Gyr,
respectively. Similarly, \citet{kilic12} and \citet{kalirai12} constrain the ages
of six kinematically-confirmed field halo white dwarfs to 11-11.5 Gyr.

Recent large scale proper motion surveys \citep{munn04,harris06,rowell11,limoges15}
provide an excellent opportunity to create accurate luminosity functions for the disk.
In addition, the proper motion survey of \citet{munn14,munn17} goes deep enough ($r=21.5$
mag) to uncover a sizable fraction of halo white dwarfs. Here we take advantage
of this latter survey to constrain, for the first time, the ages of the thin disk,
thick disk, and the halo from the same sample of white dwarfs.
In Section 2 we describe the input model white dwarf luminosity
functions, and in Section 3 we present an analysis of the local 40 pc white dwarf 
sample \citep{limoges15}. In Section 4, we analyze the
disk and halo luminosity functions from the deep proper motion survey of
\citet{munn14,munn17} and present new age constraints for the three major
components of the Galaxy. We compare these white dwarf age constraints with
previous age measurements from the literature and highlight the next generation
of surveys that will improve these age constraints in Section 5.

\section{Model Luminosity Functions}

\subsection{Thin Disk}

The canonical age estimates for the thin disk, thick disk, and halo range from about
7 Gyr for the disk to about 13 Gyr for the halo. \citet{cignoni06} and
\citet{tremblay14} find an approximately constant or a modestly rising star formation
rate history in the Galactic disk. For simplicity, we assume a constant star formation
rate for the thin disk. We generate 100 main-sequence stars at every
1 Myr timestep with masses randomly drawn between 0.8 and 8 $M_{\odot}$ from a
Salpeter mass function with exponent $\alpha=2.35$ \citep{salpeter55}. This mass
range encompasses the progenitors of all white dwarfs that would have formed in
the past 13 Gyr of Galactic history. We run the simulations until the required
age for each luminosity function is reached, thereby creating 1.3 million stars for
the 13.0 Gyr model.

We assume solar metallicity for the thin disk stars
and use the main-sequence lifetimes from the stellar evolution calculations 
by \citet{hurley00}. For the stars that evolve into a white dwarf within the age
of a given model, we use the initial-final mass relation from \citet{kalirai08}
to estimate the final masses. The model luminosity function age minus the
formation time of the main-sequence star minus the main-sequence lifetime gives
the white dwarf cooling age of each star. We then use the evolutionary models
by \citet{fontaine01} to calculate the bolometric magnitude
of each white dwarf, and bin the luminosity function with the same binning as the
observational sample studied. We create disk luminosity functions with ages ranging
from 6 to 13 Gyr, with a resolution of 0.1 Gyr.

\begin{figure}
\plotone{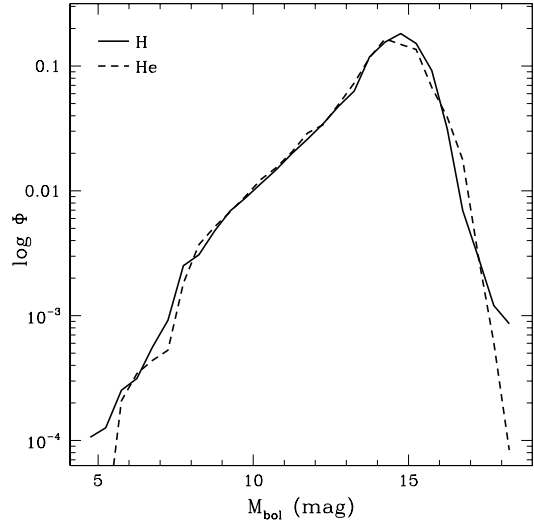}
\caption{Model luminosity functions for 10.0 Gyr old thin disk populations
of pure H (solid line) and pure He atmosphere (dashed line) white dwarfs.}
\label{fig:hhe}
\end{figure}

Figure \ref{fig:hhe} shows theoretical thin-disk luminosity functions for an
age of 10.0 Gyr for pure H (solid line) and pure He (dashed line) white dwarfs.
The two luminosity functions are remarkably similar for $M_{\rm bol}<15$ mag,
which corresponds to $T_{\rm eff}\geq 5000$ K. Both pure H and pure He
atmosphere white dwarfs with $M=0.6 M_{\odot}$ take 6.5 Gyr to reach $M_{\rm bol}=15$
mag. On the other hand, a pure H atmosphere white dwarf takes 0.7 and 0.8 Gyr longer
than a pure He atmosphere white dwarf to cool down to $M_{\rm bol}=$ 16 and 16.5 mag,
respectively. Hence, the choice of H/He atmospheres affects the faint end of the luminosity
function for $M_{\rm bol}>15$ mag and $T_{\rm eff}\leq 5000$ K.

The overall fractions of DA and DB white dwarfs
in the solar neighborhood are 80\% and 20\%, respectively \citep[e.g.][]{limoges15}.
However, the evolution of the surface composition of white dwarfs is not well
understood \citep{bergeron01}.
This is especially a problem when both He and H become invisible below 5,000 K.
Using the state-of-the-art white dwarf model atmospheres that include the
Lyman $\alpha$ red wing opacity, \citet{kowalski06} demonstrate that most or perhaps
all of the cool DC white dwarfs have hydrogen-rich atmospheres
\citep[see the discussion in][]{limoges15}.
Since these stars define the faint end of the luminosity function, and therefore the
age of a given population, we use the evolutionary models for pure H atmosphere
white dwarfs throughout our simulations. We discuss the choice of H versus He atmospheres
further in section 4.3, where we demonstrate that the addition of He atmosphere stars
($\sim$20\% He fraction) has a negligible effect on our age measurements.

\subsection{Thick Disk and Halo}

There are significant differences between the star formation history and metallicity
of the thin disk, thick disk, and the halo. 
\citet{ivezic08} used Sloan Digital Sky Survey photometry for more than 2 million
F/G stars to derive [Fe/H] = $-0.7$ and $-1.5$ for the thick disk and halo, respectively.
The thick disk is a relatively old population ($\sim$10 Gyr) with the bulk of the stars
forming over a relatively short period of time of $\sim$1 Gyr. Similarly, the Galactic
halo is also best described by a single star burst model lasting over 1 Gyr with an
age of $\sim$12 Gyr \citep{reid05}. 

To assemble the model luminosity functions for the thick disk and halo, we use a
top-hat star formation rate; we generate 1000 main-sequence stars at every 1 Myr
timestep with masses randomly drawn between 0.75 and 8 $M_{\odot}$ from a
Salpeter mass function, and run the simulations for 1 Gyr after the formation of each
population, creating 1 million stars in the process. We use the evolutionary models
by \citet{hurley00} for the thick disk and halo metallicities to
estimate the main-sequence lifetimes. We then follow the same procedure as described
in the previous section to estimate the final white dwarf mass, cooling age, and
the bolometric magnitude. We create thick disk (halo) luminosity functions with
ages ranging from 8 to 13 (15) Gyr, with a resolution of 0.1 Gyr.

There are marked differences between the model luminosity functions for the populations
with different metallicities and star formation histories. \citet{hurley00} predict
main-sequence lifetimes of 1164, 909, and 767 Myr for $2M_{\odot}$ thin disk, thick
disk, and halo stars, respectively. For an 8 Gyr old population with the star formation
histories discussed above, 44.3\% of the
thin disk stars with masses $0.8-8 M_{\odot}$ have already turned into white dwarfs,
whereas this number is significantly higher for the thick disk and the halo; 73.0\%
and 79.5\%, respectively. In addition, the thick disk and halo
luminosity functions peak at fainter magnitudes. For an 8 Gyr old population, 
76.5\% of the thin disk white dwarfs and 91.6\% of thick disk white dwarfs have 
$M_{\rm bol}>13$ mag, respectively.

\section{The 40 pc Sample}

\subsection{Previous Work}

\citet{limoges15} performed a detailed model atmosphere analysis of the local 40 pc
white dwarf sample and constructed a luminosity function based on 501 objects.
Comparing this luminosity function to the theoretical luminosity functions from
\citet{fontaine01}, \citet{limoges15} find a significant bump around 
$M_{\rm bol}\sim10$ mag and attribute this to an enhanced star formation rate about
300 Myr ago. Based on the observed cut-off in the luminosity function, they also
find a Galactic disk age of around 11 Gyr. We note that the theoretical luminosity
functions used in that work rely on simplified versions of the initial-final mass relation
and main-sequence lifetimes, specifically $M_{\rm WD} = 0.4 e^{0.125 M} M_{\odot}$ and
$t_{\rm MS} = 10 M^{-2.5}$ Gyr.

\citet{torres16} revisit the analysis of the 40 pc sample with a population synthesis
code. They draw their sample of stars from a Salpeter mass function, use the
initial-final mass relation of \citet{catalan08}, and cooling tracks of \citet{renedo10}
and \citet{althaus07}. Using the observed cut-off of the luminosity function,
they derive an age of $8.9 \pm 0.2$ Gyr and they also explain the bump around
$M_{\rm bol}=10$ mag as a burst of star formation 600 Myr ago. 
They test the reliability of their results against the assumed initial mass function
and the initial-final mass relation, and conclude that the age measurement
is insensitive to these input parameters, except for extreme
slopes for the initial-final mass relation. However, their best-fit model
significantly overpredicts the number of white dwarfs near the maximum of the
luminosity function. \citet{torres16} can explain this discrepancy with an
initial-final mass relation that has a slope 30\% larger than the observed
relation for stars more massive than $4M_{\odot}$ from \citet{catalan08}. 
However, there is no evidence for such a steep initial-final mass relation
\citep{kalirai08,williams09}, and we find this explanation unlikely.
Instead, we suggest a simpler explanation for the discrepancy;
the neglected contribution of thick disk white dwarfs to the faint
end of the luminosity function.

\begin{figure}
%\plotone{f2.ps}
\includegraphics[width=3.3in,angle=0]{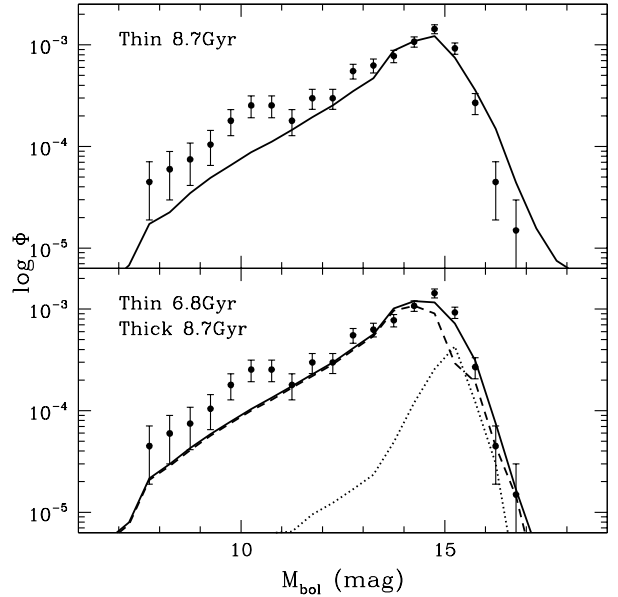}
\caption{The observed luminosity function for the local 40 pc sample of
white dwarfs \citep[points with error bars,][]{limoges15} compared to the
best-fit synthetic white dwarf luminosity function (solid lines). The top panel shows
the model fits assuming a population of 100\% thin disk stars, whereas
the bottom panel shows the fits using a composite population where
the ratio of thick disk to thin disk white dwarfs is 22\%.
Dashed and dotted lines show the contribution from the thin disk
and thick disk white dwarfs, respectively.}
\label{fig:40pc}
\end{figure}

\subsection{Thick Disk Number Density}

Studying the 3D space motions of 398 field DA white dwarfs from the Supernovae Ia
Progenitor Survey, \citet{pauli06} find that 7\% of the white dwarfs in their
sample belong to the thick disk. \citet{reid05} argue that the fraction of
thick disk white dwarfs in the local population could be as high as $\sim20$\%
given that it is an old population that leads to an enhanced number of low
luminosity objects. Since chemical-tagging of white dwarfs as thick
disk objects is not possible, identifying individual objects as thick disk members
is difficult without accurate parallaxes and constraints on their 3D space motions. 
\citet{rowell11} rely on a statistical approach to disentangle the thin disk
and thick disk luminosity functions, and find that the thick disk and halo contribute
roughly 16\% and 5\% of the local white dwarfs, respectively. Due to small number
statistics at
the faint end of their luminosity functions, they do not attempt age measurements
for the different populations, but they conclude that traditional approaches that
do not account for a significant contribution from thick disk and halo stars cannot
measure an accurate thin disk age. 

Based on star counts from the SDSS, \citet{juric08} measure a local density
ratio of $\rho_{\rm thick}/\rho_{\rm thin}=12 \pm 1$\% for main-sequence stars. 
\citet{dejong10} find a similar ratio, $15\pm4$\%, from the Sloan Extension
for Galactic Understanding and Exploration (SEGUE) survey. For an 8 Gyr old thin
disk that has continuous star formation, 44.3\% of the stars with $M=0.8-8 M_{\odot}$
turn
into white dwarfs, whereas for a 10 Gyr old thick disk that formed stars in
the first Gyr, 80.2\% are now white dwarfs. Hence, the expected number density
of thick versus thin disk white dwarfs would be $12\% \times \frac{80.2\%}{44.3\%}=21.7$\% for a kinematically
unbiased sample. However, the 40 pc sample has a lower proper motion limit of
40 mas yr$^{-1}$, which reduces the expected number of thin disk stars by $\approx$1.8\%
according to the Besan\c{c}on Galaxy model \citep{robin12}. Hence, the expected
ratio of thick versus thin disk white dwarfs is 22.1\% in the 40 pc sample.

\subsection{Thin Disk and Thick Disk Ages}

Figure \ref{fig:40pc} shows the 40 pc white dwarf luminosity function from
\citet{limoges15} kindly made available to us by P. Bergeron. To avoid the bump around $M_{\rm bol} = 10$ mag due to the enhanced
star formation rate in the past 600 Myr \citep{torres16}, we only use the stars with $M_{\rm bol}>11$ mag
in our fits. Since we are mainly interested in the age constraints from the faint
end of the luminosity function, we do not try to model this bump, as it has no
effect on our age constraints.
The top panel shows our fits assuming a population of thin disk stars
only. The best fit thin disk age is $8.7 \pm 0.1$ Gyr, which is consistent with
$8.9 \pm 0.2$ Gyr as measured by \citet{torres16}. However, this age measurement
is clearly wrong since it ignores the contribution of thick disk stars and it assumes
solar metallicity for all objects in the sample.

We show the best-fitting thin disk + thick disk composite luminosity function,
as well as the contributions from both populations in the bottom panel. For
$\rho_{\rm thick,WD}/\rho_{\rm thin,WD}=$ 22\%, the best-fitting model has
ages of $6.8 \pm 0.2$ and  $8.7 \pm 0.1$ Gyr for the thin disk and thick disk,
respectively. The composite thin+thick disk model is a significantly better fit
than a thin-disk only model ($\chi^2_{red}=$ 1.6 versus 3.5). In addition,
the age constraints are insensitive to the assumed thick/thin disk fraction
as long as this fraction is above a few per cent.

We note that the uncertainties in the age estimate comes from
a Monte Carlo analysis where we replace the measured space density $\phi$ with
$\phi + g \delta\phi$, where $\delta\phi$ is the error in space density
and $g$ is a Gaussian deviate with zero mean and unit variance. For each of
1,000 sets of modified luminosity functions, we find the model that provides
the lowest $\chi^2$ fit, and we take the range in age that encompasses 68\% of
the probability distribution function as the $1\sigma$ uncertainties.

To test the sensitivity of our age measurements to the specific binning used by 
\citet{limoges15}, we create a new luminosity function for the 40 pc sample where
the bin centers are shifted by 0.25 mag. Using the stars with $M_{\rm bol}>11$ mag
and assuming a thick/thin disk ratio of 0.22, we derive ages of $7.0^{+0.1}_{-0.2}$ Gyr
and  $8.6 \pm 0.1$ Gyr for the thin disk and thick disk, respectively. Hence, the
derived ages are not sensitive to the binning used in the luminosity
function, and they are constrained to be 6.8-7.0 Gyr and 8.6-8.7 Gyr for the thin disk
and thick disk, respectively.

\section{The Deep Proper Motion Survey Sample}

\subsection{The Disk Luminosity Function}

\citet{munn14} presented a $\sim$3,000 square degree deep proper motion survey reaching
a limiting magnitude of $r=21-22$ mag. \citet{munn17} identify 8472 white dwarf
candidates in this survey, and use 2839 stars with $M_{\rm bol}=5.5-17$ mag to
create a disk white dwarf luminosity function. To avoid
contamination from subdwarfs (at the low velocity end), they limit their disk
sample to objects with $v_{\rm tan} =$ 40-120 km s$^{-1}$. The 40 pc white dwarf
sample contains 84 objects with $M_{\rm bol}>15$ mag, whereas the \citet{munn17}
disk luminosity function has 311 objects in the same magnitude range. Hence,
the latter survey significantly increases the number of stars beyond the turnover
in the disk luminosity function, and it provides an excellent opportunity to disentangle
the thin disk and thick disk luminosity functions.

\begin{figure}
%\plotone{f3.ps}
\includegraphics[width=3.3in,angle=0]{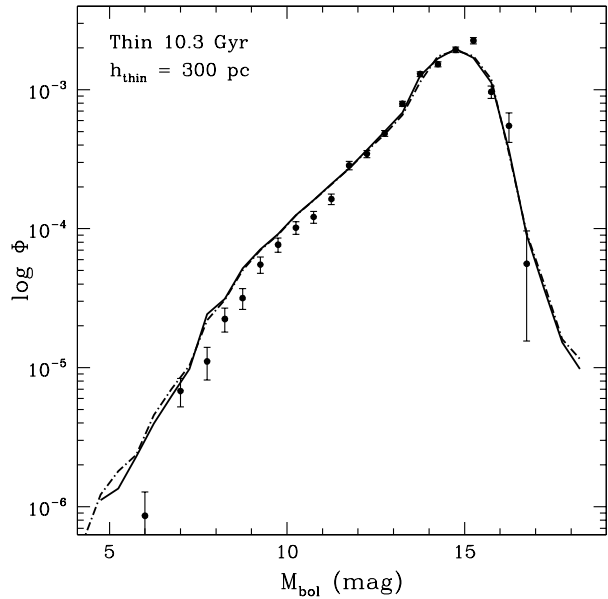}
\caption{White dwarf luminosity function from the deep proper motion survey
\citep[points with error bars,][]{munn17} assuming thin and
thick disk scale heights of 300 and 900 pc, respectively. 
The solid and dot-dashed lines show
the best-fitting model luminosity functions for a 100\% thin disk population using
the \citet{kalirai08} and \citet{williams09} initial-final mass relations, respectively.}
\label{fig:disk1}
\end{figure}

Figure \ref{fig:disk1} presents the disk luminosity function from \citet{munn17} using
the preferred disk model of \citet{juric08} with thin and thick disk scale heights
of 300 and 900 pc, respectively, and a thick to thin disk ratio of 12\%.
As in the analysis of the 40 pc luminosity function, we only use the stars with
$M_{\rm bol}>11$ mag in our fits. Assuming a population of thin disk stars only,
the best fit disk age is $10.3_{-0.1}^{+0.3}$ Gyr.
The best-fitting disk model underpredicts the peak of the luminosity function, which
is likely due to the missing contribution of thick disk stars in these fits. However,
what is striking is that the models also overpredict the number of stars brighter
than $M_{\rm bol}=11$ mag. The dashed-dotted line shows the same
model luminosity function using the \citet[][instead of Kalirai et al. 2008]{williams09}
initial-final mass relation. The difference between these two model luminosity functions
is negligible. Hence, the choice of initial-final mass relation cannot explain the
discrepancy between the observed and theoretical luminosity functions.

\citet{harris06} chose a disk scale height of 250 pc in their analysis, and they
demonstrate (see their Figure 6) that changing the scale height from 200 to 350 pc
has a significant impact on the shape of the luminosity function at the bright end.
However, a constant scale height for the entire disk is an over-simplification.
In what follows, we develop a more realistic scaleheight correction for the disk
luminosity function.

\citet{bonatto06} demonstrate that the scale height increases from $48 \pm 3$ pc
for open clusters younger than 200 Myr to $150 \pm 27$ pc for clusters with ages up
to 1 Gyr. Studying a larger sample of open clusters, \citet{buckner14} and
\citet{joshi16} show that this trend continues for older clusters; they find
a scale height $h>300$ pc for 2.5 Gyr and older clusters.

\citet{bovy12} use SEGUE G dwarfs to define mono-abundance populations, and fit
each population as a single exponential disk. The scale height for these populations
varies smoothly from around 200 pc for the most metal rich population to about 900 pc
for the most metal poor population. They interpret the decreasing metallicity as an
indication of increasing age, and their 900 pc limit is consistent with the thick
disk scale height found by \citet{juric08}.

\begin{deluxetable*}{rcccc}
\tabletypesize{\scriptsize}
\tablecolumns{11}
\tablewidth{0pt}
\tablecaption{White Dwarf Luminosity Functions Using Different Scale Heights}
\tablehead{
\colhead{$M_{\rm bol}$}&
\colhead{$\Phi_{\rm 200-900}$}&
\colhead{$\Phi_{\rm 200-700}$}&
\colhead{$\Phi_{\rm 200-500}$}&
\colhead{$\Phi_{\rm Munn et al. (2017)}$}\\
(mag) & (pc$^{-3} M_{\rm bol}^{-1})$ & (pc$^{-3} M_{\rm bol}^{-1})$ & (pc$^{-3} M_{\rm bol}^{-1})$ & (pc$^{-3} M_{\rm bol}^{-1})$
}
\startdata
6.00  & 1.337e-06 $\pm$ 6.487e-07 & 1.366e-06 $\pm$ 6.628e-07 & 1.396e-06 $\pm$ 6.775e-07 & 8.196e-07 $\pm$ 3.945e-07\\
7.00  & 9.838e-06 $\pm$ 2.284e-06 & 1.018e-05 $\pm$ 2.364e-06 & 1.054e-05 $\pm$ 2.450e-06 & 6.888e-06 $\pm$ 1.536e-06\\
7.75  & 1.509e-05 $\pm$ 3.998e-06 & 1.573e-05 $\pm$ 4.169e-06 & 1.644e-05 $\pm$ 4.358e-06 & 1.219e-05 $\pm$ 3.006e-06\\
8.25  & 3.104e-05 $\pm$ 6.105e-06 & 3.212e-05 $\pm$ 6.319e-06 & 3.330e-05 $\pm$ 6.551e-06 & 2.236e-05 $\pm$ 4.313e-06\\
8.75  & 3.768e-05 $\pm$ 6.498e-06 & 4.020e-05 $\pm$ 6.932e-06 & 4.328e-05 $\pm$ 7.464e-06 & 3.656e-05 $\pm$ 5.699e-06\\
9.25  & 6.406e-05 $\pm$ 8.585e-06 & 6.845e-05 $\pm$ 9.174e-06 & 7.394e-05 $\pm$ 9.910e-06 & 5.464e-05 $\pm$ 7.222e-06\\
9.75  & 8.614e-05 $\pm$ 9.988e-06 & 9.231e-05 $\pm$ 1.070e-05 & 1.002e-04 $\pm$ 1.162e-05 & 8.218e-05 $\pm$ 9.003e-06\\
10.25  & 1.167e-04 $\pm$ 1.208e-05 & 1.243e-04 $\pm$ 1.287e-05 & 1.338e-04 $\pm$ 1.385e-05 & 1.097e-04 $\pm$ 1.071e-05\\
10.75  & 1.363e-04 $\pm$ 1.336e-05 & 1.452e-04 $\pm$ 1.424e-05 & 1.567e-04 $\pm$ 1.536e-05 & 1.237e-04 $\pm$ 1.183e-05\\
11.25  & 1.802e-04 $\pm$ 1.580e-05 & 1.919e-04 $\pm$ 1.682e-05 & 2.069e-04 $\pm$ 1.813e-05 & 1.680e-04 $\pm$ 1.431e-05\\
11.75  & 3.042e-04 $\pm$ 2.082e-05 & 3.241e-04 $\pm$ 2.219e-05 & 3.504e-04 $\pm$ 2.400e-05 & 2.972e-04 $\pm$ 1.964e-05\\
12.25  & 3.615e-04 $\pm$ 2.123e-05 & 3.842e-04 $\pm$ 2.256e-05 & 4.144e-04 $\pm$ 2.434e-05 & 3.556e-04 $\pm$ 2.029e-05\\
12.75  & 4.882e-04 $\pm$ 2.352e-05 & 5.192e-04 $\pm$ 2.501e-05 & 5.614e-04 $\pm$ 2.704e-05 & 5.044e-04 $\pm$ 2.353e-05\\
13.25  & 7.675e-04 $\pm$ 3.213e-05 & 8.154e-04 $\pm$ 3.413e-05 & 8.817e-04 $\pm$ 3.691e-05 & 8.164e-04 $\pm$ 3.334e-05\\
13.75  & 1.166e-03 $\pm$ 4.525e-05 & 1.237e-03 $\pm$ 4.801e-05 & 1.345e-03 $\pm$ 5.220e-05 & 1.345e-03 $\pm$ 5.072e-05\\
14.25  & 1.296e-03 $\pm$ 5.325e-05 & 1.367e-03 $\pm$ 5.612e-05 & 1.482e-03 $\pm$ 6.086e-05 & 1.565e-03 $\pm$ 6.320e-05\\
14.75  & 1.620e-03 $\pm$ 7.352e-05 & 1.691e-03 $\pm$ 7.674e-05 & 1.815e-03 $\pm$ 8.232e-05 & 1.961e-03 $\pm$ 8.793e-05\\
15.25  & 1.881e-03 $\pm$ 9.947e-05 & 1.947e-03 $\pm$ 1.030e-04 & 2.066e-03 $\pm$ 1.093e-04 & 2.292e-03 $\pm$ 1.197e-04\\
15.75  & 8.211e-04 $\pm$ 8.394e-05 & 8.453e-04 $\pm$ 8.641e-05 & 8.888e-04 $\pm$ 9.087e-05 & 9.945e-04 $\pm$ 9.927e-05\\
16.25  & 4.860e-04 $\pm$ 1.171e-04 & 4.981e-04 $\pm$ 1.199e-04 & 5.194e-04 $\pm$ 1.247e-04 & 5.568e-04 $\pm$ 1.315e-04\\
16.75  & 4.990e-05 $\pm$ 3.602e-05 & 5.124e-05 $\pm$ 3.698e-05 & 5.353e-05 $\pm$ 3.863e-05 & 5.591e-05 $\pm$ 4.034e-05
\enddata
\end{deluxetable*}

\begin{figure}
%\plotone{f4.ps}
\includegraphics[width=3.3in,angle=0]{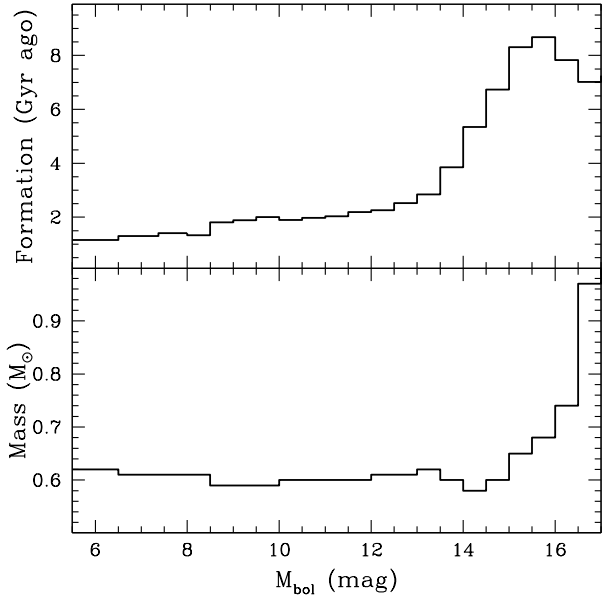}
\caption{The median formation times for the progenitor stars and the median masses for the white dwarfs
in each magnitude bin of the synthetic luminosity function for a 10.0 Gyr old
thin disk population.}
\label{fig:formation}
\end{figure}

Figure \ref{fig:formation} shows the median formation times of the
progenitor main-sequence stars and median masses for
the white dwarfs in each magnitude bin of a 10.0 Gyr old thin disk luminosity function.
Based on our synthetic luminosity functions, the median formation times
range from 1.2 Gyr at $M_{\rm bol}=6$ mag to 8.3 Gyr
at $M_{\rm bol}=15.25$ mag. Thus, instead of a constant disk scale height, we let
the scale height vary linearly between 200 pc for age = 1 Gyr to 900 pc for age =10 Gyr.

\begin{figure}
%\plotone{f5.ps}
\includegraphics[width=3.3in,angle=0]{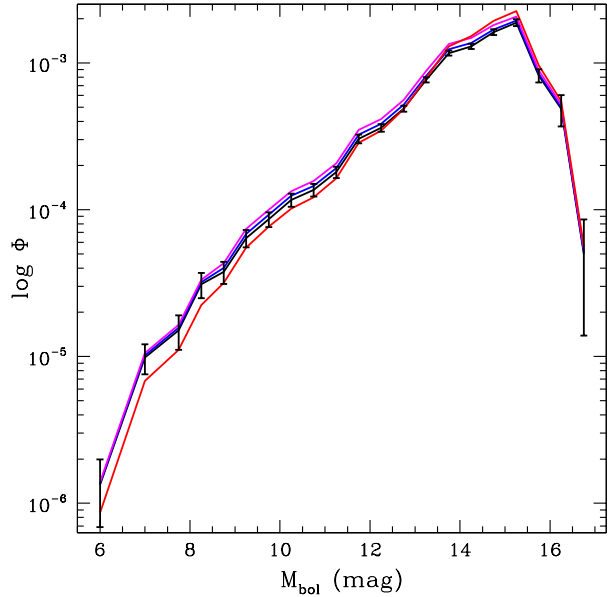}
\caption{Luminosity functions using the \citet{juric08}
disk model (red), and models with scale heights that vary linearly
with age from 200 pc at 1 Gyr to 500 (magenta), 700 (blue), and 900 pc (black) at 10 Gyr.}
\label{fig:diskb}
\end{figure}

Table 1 and Figure \ref{fig:diskb} compare the luminosity function with that
age/scale-height relation applied to the median ages that we obtain. For comparison,
we also overplot the preferred model from \citet[][where $h_{\rm thin}=300$ pc]{munn17},
along with luminosity functions where the upper limit in scale height is 700 and
500 pc. The variable scale-height luminosity functions shown in Fig. \ref{fig:diskb} and Table 1 are
consistent within the errors, but in the $M_{\rm bol}<11$ mag range they have a
significantly larger space density than the luminosity function assuming a
constant thin disk scale height of 300 pc.
Hence, the discrepancy between the observed and model luminosity
functions in Figure \ref{fig:disk1} is likely due to our previous assumption
of a constant scale height for the disk.

\subsection{Thin Disk and Thick Disk Ages}

Following the discussion from Section 3.2, the expected white dwarf number density
is $\rho_{\rm thick}/\rho_{\rm thin}=$ 21.7\% for a kinematically unbiased population.
\citet{munn17} correct the disk luminosity function for missing objects with
$V_{\rm tan}<40$ or $>120$ km s$^{-1}$ using the modified maximum survey volume of
\citet{lam15} as the density estimator. However, they treat the disk objects as a single
population, and sum the thin and thick disk profiles of \citet{juric08} as a single disk
density profile and use the \citet{fuchs09} results to model the kinematics of the disk. Hence,
they correct for the overall number of disk objects in their survey, but these corrections
do not include the change in the relative numbers of thin and thick disk objects due to the
$V_{\rm tan}=$ 40-120 km s$^{-1}$ cut.

The median Galactic latitude of the fields observed by \citet{munn17} is
$|b|=50^{\circ}$, with the range 34$^{\circ}$-63$^{\circ}$ containing 68\% of the
observed  fields.
Based on the Besan\c{c}on Galaxy model for $|b|=50^{\circ}$, we expect 59.1\% and
33.7\% of thin disk and thick disk stars to be lost due to the 40-120 km s$^{-1}$ tangential
velocity cut. Hence, the expected ratio of thick versus thin disk white dwarfs
in the deep proper motion survey disk luminosity function is 35.2\%. Using the
$|b|=34^{\circ}$-63$^{\circ}$ range, the expected ratio is between 32.2 and 41.6\%,
with an additional 3-4\% uncertainty due to the density uncertainty estimates from
\citet{juric08}.

\begin{figure}[b]
%\plotone{f6.ps}
\includegraphics[width=3.3in,angle=0]{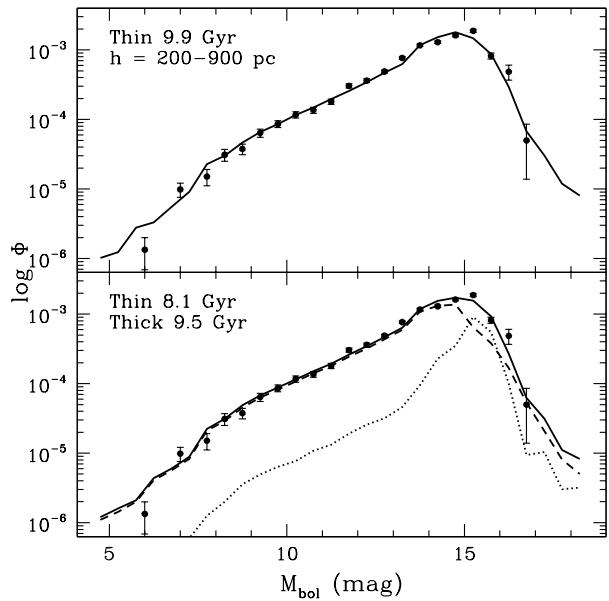}
\caption{The white dwarf luminosity function from the deep proper motion survey
\citep[points with error bars,][]{munn17} using a disk scale height range of
200-900 pc. The top panel shows the model fits assuming a population of 100\% thin
disk stars, whereas the bottom panel shows the fits using a composite population where
the ratio of thick disk to thin disk white dwarfs is 35\%.
Dashed and dotted lines show the contribution from the thin disk
and thick disk white dwarfs, respectively.}
\label{fig:disk2}
\end{figure}

Figure \ref{fig:disk2} shows the white dwarf luminosity function from the deep proper
motion survey using a variable disk scale height of 200-900 pc.
The top panel shows our fits assuming a population of thin disk stars
only, which leads to an age estimate of 9.9 Gyr. Interestingly, the observed
and best-fitting model luminosity functions agree within $2.3\sigma$ in the 
$M_{\rm bol}=$ 7-13 mag range. Accounting for a disk scale
height that increases with age yields luminosity functions consistent with a
constant star formation rate in the past $\sim$2.5 Gyr.

The bottom panel in Figure \ref{fig:disk2} shows the best-fitting thin disk + thick
disk composite luminosity function for $\rho_{\rm thick,WD}/\rho_{\rm thin,WD}=$35\%.
The best-fitting model has ages of $8.1_{-0.1}^{+0.2}$ Gyr and  $9.5 \pm 0.2$ Gyr
for the thin disk and thick disk, respectively. The composite thin+thick disk model
matches the peak of the luminosity function relatively well.

\begin{figure}[b]
\plotone{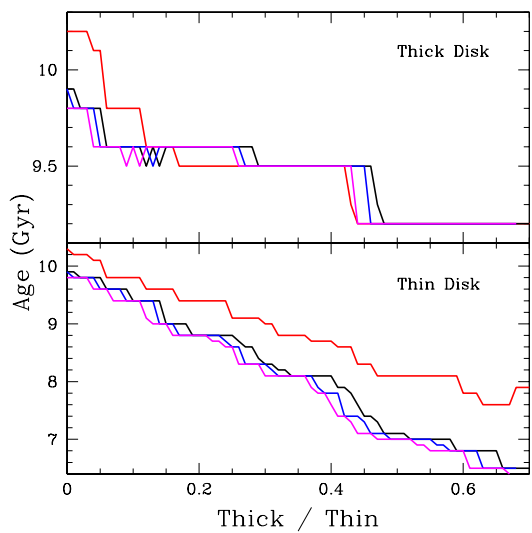}
\caption{Thin disk and thick disk age constraints as a function
of the ratio of thick disk to thin disk white dwarfs. The black, blue, magenta,
and red lines show the constraints from luminosity functions using the scale
heights of 200-900 pc \citep{bovy12}, 200-700 pc, 200-500 pc, and 300 pc
\citep{juric08}, respectively.}
\label{fig:fraction}
\end{figure}

Figure \ref{fig:fraction} shows age constraints on the thin and thick disk
as a function of the number density of thick versus thin disk white dwarfs. 
We use the same color-scheme as in Figure \ref{fig:diskb} to indicate the models
with different scale heights. This figure demonstrates that the thick disk age
is constrained relatively well to $9.5 \pm 0.2$ Gyr, even in the luminosity
function that uses a constant disk scale height of 300 pc. This is because the
thick disk is an old population, and most of the age sensitivity comes from the
observed cut-off in the luminosity function, which does not vary considerably
between the different luminosity functions presented in Figure \ref{fig:diskb}.
On the other hand, the thin disk age measurement heavily depends on the shape
of the broad peak observed above $M_{\rm bol}=13$ mag and it is sensitive to
the assumed thick/thin disk population fraction.
For a thick/thin ratio of 35\%, the age of the thin disk is 8.1 Gyr for models
that use a variable scale height. However, the thin disk age varies from 7.4 to
8.2 Gyr for a thick/thin ratio in the range 32\%-42\%.
Hence, based on the \citet{munn17} disk luminosity function, we adopt an age of
7.4-8.2 Gyr for the thin disk.

\subsection{Thin Disk and Thick Disk Ages: Caveats}

There are two caveats in our analysis of the disk ages. The first one is the 
assumed fractions of hydrogen versus helium atmosphere white dwarfs in our 
synthetic luminosity functions. Based on the analysis of \citet{kowalski06},
we assumed a 100\% fraction of H atmosphere white dwarfs in our analysis.
\citet{torres16} use fractions of 80\% DA and 20\% DB in their analysis of
the 40 pc sample. To explore the effects of DB white dwarfs on our age measurements,
we repeat our analysis of the disk luminosity function assuming a DB
white dwarf fraction of 20\%.

\begin{figure}
%\plotone{f8.ps}
\includegraphics[width=3.3in,angle=0]{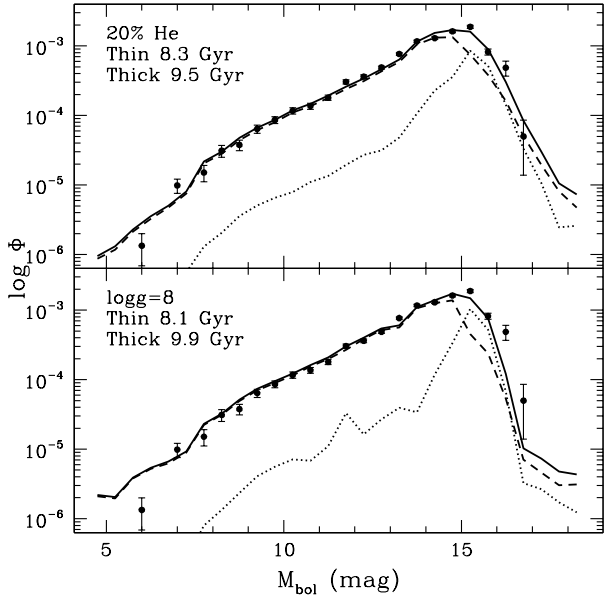}
\caption{Fits to the disk luminosity function assuming that 20\% of the white
dwarfs have pure Helium atmospheres (top panel), or 100\% of them have $\log{g}=8$
(bottom panel). Dashed and dotted lines show the contribution from the thin disk
and thick disk white dwarfs, assuming a 35\% thick/thin ratio, respectively.}
\label{fig:helium}
\end{figure}

Figure \ref{fig:helium} shows these fits using
a 20\% pure He atmosphere fraction for both the thin disk and thick disk populations
(top panel). For a ratio of $\rho_{\rm thick}/\rho_{\rm thin}=35$\%, the best-fitting
model has ages of $8.3_{-0.2}^{+0.3}$ Gyr and  $9.5_{-0.3}^{+0.1}$ Gyr
for the thin disk and thick disk, respectively. These are consistent with our previous
disk age estimates within 1$\sigma$. Hence, the choice of DA versus DB atmospheres
has a negligible effect on our age measurements for a DB white dwarf fraction of $\sim$20\%.

Due to the lack of parallax measurements, \citet{munn17} had to assume $\log{g}=8$
for all stars in their sample. This is the second caveat in our analysis as our synthetic
luminosity functions draw stars with random initial masses from a Salpeter mass function
and use the initial-final mass relation to estimate the white dwarf masses. The bottom
panel in Figure \ref{fig:formation} shows the median masses for the white dwarfs in each
magnitude bin of the synthetic luminosity function for a 10 Gyr old thin disk population.
The median mass ranges between 0.58 and 0.62 $M_{\odot}$ for $M_{\rm bol}<15$ mag. Hence,
the assumption of $\log{g}=8$ is appropriate for getting the overall shape of the luminosity
function right. However, the median mass increases to $\approx 0.7 M_{\odot}$ at
$M_{\rm bol}=16$ mag, which has implications for the age measurements from the faint end
of the lumionsity function.

\citet{munn17} fit the SDSS photometry of each target with the $\log{g}=8$ white
dwarf models to derive its $T_{\rm eff}$ and bolometric magnitude.
To explore the effects of the $\log{g}=8$ assumption on our age estimates, we create synthetic
luminosity functions where we use the cooling age of each white dwarf to estimate its
$T_{\rm eff}$, and calculate its bolometric luminosity at that temperature assuming
$\log{g}=8$. The bottom panel in Figure \ref{fig:helium} shows these fits for $\log{g}=8$
white dwarfs. The best-fitting model has ages of $8.1_{-0.1}^{+0.2}$ Gyr and $9.9 \pm 0.1$ Gyr
for the thin disk and thick disk, respectively. 

The thin disk age is essentially unchanged
from our previous estimates, as the peak of the thin disk luminosity function is below
$M_{\rm bol}=15$ mag, where the median mass in our synthetic luminosity functions is around
0.6 $M_{\odot}$. However, the thick disk age is sensitive to the magnitude bins where the
median mass is $\approx0.7 M_{\odot}$. Based on the \citet{fontaine01} cooling models,
a 4,000 K, 0.7 $M_{\odot}$ white dwarf has a bolometric magnitude of 16.1,
whereas a 0.6 $M_{\odot}$ white dwarf at the same temperature has $M_{\rm bol}=15.9$ mag.
Hence, the assumption of $\log{g}=8$ underestimates the bolometric magnitudes of the white
dwarfs at the faint end of the luminosity function. This requires a larger age for our
targets to reach the observed cut-off in magnitude. Given this systematic uncertainty in
age, we adopt a best-fit age of 9.5-9.9 Gyr for the thick disk.

\subsection{The Halo Luminosity Function}

Previous efforts to create halo white dwarf luminosity functions using
field stars have suffered from small number statistics. For example,
\citet{harris06} identify 18 stars with $v_{\rm tan}>200$ km s$^{-1}$ as
halo white dwarf candidates, including two stars with $M_{\rm bol}>14$ mag.
\citet{rowell11} find 93 stars with $v_{\rm tan}>200$ km s$^{-1}$ and UVW
space velocities that are consistent with a non-rotating population (the spheroid).
Their luminosity function includes objects as faint as $M_{\rm bol}=15$ mag,
though the faint end of their luminosity function has relatively large
error bars due to small numbers of stars in those magnitude bins.

\begin{figure}
%\plotone{f9.ps}
\includegraphics[width=3.3in,angle=0]{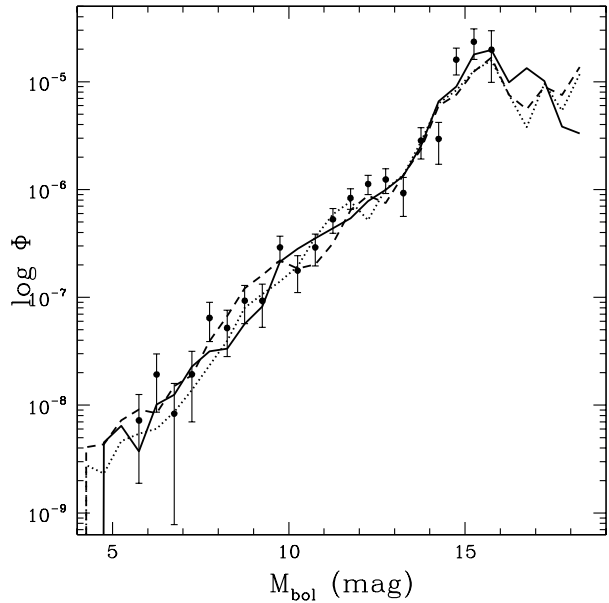}
\caption{\citet{munn17} luminosity function for $v_{\rm tan}=$ 200-500 km s$^{-1}$
halo white dwarf sample. Solid, dashed, and dotted lines show model
luminosity functions for 12.5, 13.9, and 15.0 Gyr old halo samples, respectively.
This luminosity function implies a halo age of $12.5^{+1.4}_{-3.4}$ Gyr.}
\label{fig:halo}
\end{figure}

\citet{munn17} use 135 stars with $M_{\rm bol}=5.5-17$ mag to create a halo
white dwarf luminosity function. To minimize the contamination from the large
number of disk objects, they limit their halo sample to objects with
$v_{\rm tan} =$ 200-500 km s$^{-1}$. Figure \ref{fig:halo} shows this
luminosity function, which contains 21 stars with $M_{\rm bol}=$ 14-16 mag. 
The halo white dwarf luminosity function rises nearly monotonically to
$M_{\rm bol}=$ 16 mag, though the turnover at the faint end remains undetected. 
Fitting this luminosity function with our models based on a Salpeter initial
mass function, star formation in the first Gyr after its formation, \citet{hurley00}
main-sequence lifetimes for metal-poor stars, \citet{kalirai08} initial-final mass 
relation, \citet{fontaine01} evolutionary models, and based on 10,000 Monte-Carlo
simulations, we find that the halo age is constrained to $\geq8.1$ Gyr at the
99.7\% confidence level (3$\sigma$). The best-fit halo age is $12.5^{+1.4}_{-3.4}$ Gyr,
and the 1$\sigma$ upper limit of 13.9 Gyr is consistent with the latest constraints
on the age of the Universe \citep{planck16}. However, this luminosity function
is not deep enough to show a cutoff at the faint end,
and is consistent with an arbitrarily large halo age of 15.0 Gyr
(the upper age limit of our model luminosity functions) within 2$\sigma$.

Figure \ref{fig:halo} shows the best-fitting model (12.5 Gyr, solid line)
luminosity function, as well as the models for 13.9 Gyr (dashed) and 15.0 Gyr
(dotted line) old halo. Our halo age estimate is consistent with the age
measurements for the inner halo from the field halo white dwarfs of
\citet[][$11.4\pm0.7$ Gyr]{kalirai12}, \citet[][11-11.5 Gyr]{kilic12}, and
\citet[][$11.67^{+1.02}_{-0.96}$ Gyr]{si17}, as well
as the ages of the globular cluster white dwarf sequences of M4, NGC 6397, and 47 Tuc,
which span a range of 9.9 to 11.5 Gyr. However, better age constraints for the
inner halo would require a luminosity function that goes at least a magnitude deeper
than the current surveys.

\section{Conclusions}

We present an analysis of the white dwarf luminosity functions from the local 40 pc
sample and the deep proper motion survey of \citet{munn17}. We demonstrate that
both samples have significant numbers of thick disk stars that define the faint
end of their luminosity functions. Simultaneously fitting for both disk components,
we constrain the ages of the thin disk and thick disk to be 6.8-7.0 Gyr and 8.7 $\pm$
0.1 Gyr from the local sample, and 7.4-8.2 Gyr and 9.5-9.9 Gyr from the
\citet{munn17} sample of white dwarfs, respectively.  

\begin{figure}
%\plotone{f10.ps}
\includegraphics[width=3.3in,angle=0]{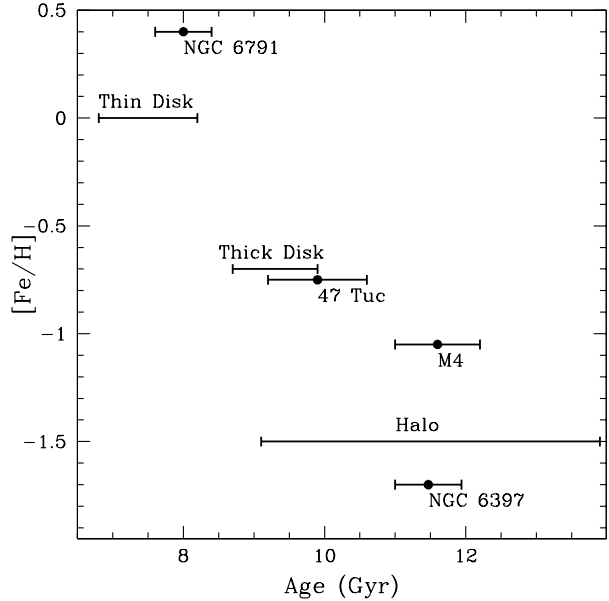}
\caption{Age-Metallicity relation based on the white dwarf luminosity functions
for the open cluster NGC 6791, globular clusters 47 Tuc, M4, and NGC 6397
\citep[][and references therein]{hansen13}, and field thin disk, thick disk, and
halo stars from this study. The error bars cover the age ranges estimated from
both the 40 pc local sample and the \citet{munn17} deep proper motion survey sample.}
\label{fig:ages}
\end{figure}

Figure \ref{fig:ages} presents the age-metallicity relation based on the
white dwarf luminosity functions for four clusters studied with the {\em Hubble Space
Telescope} \citep{hansen13}. This figure also includes our results
from field white dwarfs. Our thin disk age estimate of
6.8-8.2 Gyr is larger than the ages of the oldest, solar-metallicity open clusters 
observed, e.g. M67
\citep[4 Gyr,][]{sandquist04,vonhippel05} and NGC 188 \citep[6.2 Gyr,][]{meibom09}. However, this
is not surprising, since older clusters are tidally disrupted over the lifetime of the
disk \citep{soderblom10}. In addition, this age is in excellent agreement
with the $7.3 \pm 1.5$ Gyr age estimate from the \citet{liebert88} white dwarf luminosity
function by \citet{hansen02}, as well as the ages of $8.0 \pm 0.4$ Gyr for the
metal-rich cluster NGC 6791 \citep{garciaberro10} and 7.5-7.9 $\pm$ 0.7 Gyr for
the turn-off field stars and subgiants with parallax measurements \citep{liu00,sandage03}.
Our thick disk age measurements from the 40 pc and the \citet{munn17} sample differ
significantly, but the latter survey includes white dwarfs with estimated distances
of up to $\sim1$ kpc. There is a well established trend between the age of a
population and its disk scale height \citep[e.g.,][]{bovy12}, hence the relatively
younger ages estimated from the local 40 pc sample are not surprising. 
The observed age range of 8.7-9.9 Gyr for thick disk white dwarfs is in excellent agreement
with the relatively metal-rich globular cluster 47 Tuc \citep{hansen13}. 

Perhaps the most important result of this analysis is not the absolute age measurements,
but instead the demonstration of age differences between the different kinematic
populations of white dwarfs. For the 40 pc local sample, we measure relative age
differences of 1.9 $\pm$ 0.2 Gyr and 1.6$^{+0.2}_{-0.1}$ Gyr between the thin disk
and thick disk populations based on the original luminosity function presented by
\citet{limoges15} and the same luminosity function with the shifted bin centers (see \S 3.3).
Similarly, our analysis of the disk luminosity function from the deep proper motion
survey implies a relative age difference of 1.4$^{+0.1}_{-0.3}$ Gyr.
Combining the results for the local 40 pc sample and the \citet{munn17} proper motion sample,
we constrain the time between the onset of star formation in the thick disk and thin
disk to be 1.6$^{+0.3}_{-0.4}$ Gyr. This is similar to, though twice as precise as,
the relative age difference of $1.9 \pm 0.8$ Gyr between 47 Tuc and NGC 6791.

Accurate parallaxes from the Gaia mission will help in constraining the faint
end of the disk luminosity function. In addition, where radial velocity measurements are also
available, Gaia data will help separate thin and thick disk objects based on their 3D
space velocities. However, the peak of the thick disk luminosity function is beyond
$M_{\rm bol}=15$ mag and $T_{\rm eff}=$ 5000 K, below which H$\alpha$ disappears.
Hence, it is likely that future studies of the Gaia disk luminosity function will
have to rely on a method similar to ours \citep[or to][]{rowell11,lam17} in disentangling
the thin disk and thick disk luminosity functions.

Both \citet{limoges15} and \citet{torres16} find evidence of an enhanced star formation rate 
in the past 300-600 Myr in the local 40 pc white dwarf sample. On the other hand,
we do not find any evidence of a deviation from a constant star formation rate in the
past 2.5 Gyr in the \citet{munn17} disk luminosity function, which samples a more
distant population of white dwarfs. Most molecular clouds and star formation
are constrained close to the Galactic plane. The scale height for open clusters younger
than 200 Myr is only 48 pc \citep{bonatto06}. Hence, we conclude that the enhanced star formation rate 
observed in the 40 pc local sample is only relevant for the immediate
Solar neighborhood. Gaia parallaxes will enable us to create disk luminosity functions
as a function of distance and study their star formation histories.

The faint end of the luminosity function for nearby halo white dwarfs is not
constrained reliably, leading to an age estimate of $12.5^{+1.4}_{-3.4}$ Gyr.
Large scale surveys that reach at least 1 mag deeper than
the \citet{munn14} proper motion survey will be useful for better constraining
the halo luminosity function. The  Large Synoptic Survey Telescope (LSST)
will image 13 million white dwarfs down to $r=24.5$ mag. \citet{munn17} measure
a space density of 1/157 for halo/disk white dwarfs. Even with such a small density,
the LSST will identify $\sim10^5$ halo white dwarfs, which will enable precise
age estimates for the inner halo.

\acknowledgements
We gratefully acknowledge the support of the NSF and NASA under
grants AST-0607480, AST-0602288, AST-1312678, and NNX14AF65G.

\end{document}